\documentclass[letterpaper, 10 pt, conference]{ieeeconf}  % Comment this line out
\IEEEoverridecommandlockouts                              % This command is only
\usepackage[utf8]{inputenc}
\usepackage[T1]{fontenc}
\usepackage{url}
\usepackage{hyperref}
\usepackage{graphicx}
\usepackage{amsmath}

\title{\LARGE \bf
Respiratory Inhaler Sound Event Classification Using Self-Supervised Learning
}

\author{Davoud Shariat Panah$^{1}$, Alessandro N Franciosi$^{2}$, Cormac McCarthy$^{3}$, and Andrew Hines$^{4}$% <-this % stops a space
\thanks{$^{1}$D. Shariat Panah is with the School of Computer Science, University College Dublin, Ireland;
        {\tt\small davoud.shariatpanah at ucd.ie}}%
\thanks{$^{2}$A. Franciosi is with the School of Medicine, University College Dublin, Ireland, and Department of Respiratory Medicine, St. Vincent's University Hospital, Dublin, Ireland;
        {\tt\small alessandro.franciosi at ucd.ie}}%
\thanks{$^{3}$C. McCarthy is with the School of Medicine, University College Dublin, Ireland, and Department of Respiratory Medicine, St. Vincent's University Hospital, Dublin, Ireland;
        {\tt\small cormac.mccarthy at ucd.ie}}%
\thanks{$^{4}$A. Hines is with the School of Computer Science, University College Dublin, Ireland;
        {\tt\small andrew.hines at ucd.ie}}%
}

\begin{document}

\maketitle
\thispagestyle{empty}
\pagestyle{empty}

%%%%%%%%%%%%%%%%%%%%%%%%%%%%%%%%%%%%%%%%%%%%%%%%%%%%%%%%%%%%%%%%%%%%%%%%%%%%%%%%
\begin{abstract}

Asthma is a chronic respiratory condition that affects millions of people worldwide. While this condition can be managed by administering controller medications through handheld inhalers, clinical studies have shown low adherence to the correct inhaler usage technique. Consequently, many patients may not receive the full benefit of their medication. Automated classification of inhaler sounds has recently been studied to assess medication adherence. However, the existing classification models were typically trained using data from specific inhaler types, and their ability to generalize to sounds from different inhalers remains unexplored. In this study, we adapted the wav2vec~2.0 self-supervised learning model for inhaler sound classification by pre-training and fine-tuning this model on inhaler sounds. The proposed model shows a balanced accuracy of 98\% on a dataset collected using a dry powder inhaler and smartwatch device. The results also demonstrate that re-finetuning this model on minimal data from a target inhaler is a promising approach to adapting a generic inhaler sound classification model to a different inhaler device and audio capture hardware. This is the first study in the field to demonstrate the potential of smartwatches as assistive technologies for the personalized monitoring of inhaler adherence using machine learning models.

\indent \textit{Clinical relevance}— Accurate classification of inhaler sounds, which can be captured using consumer devices, is an essential step toward developing personalized systems for assessing inhaler compliance which can improve patient outcomes by ensuring proper medication delivery and effectiveness. Individual monitoring of inhaler adherence could also play a key role in evaluating medication efficacy in future personalized medicine clinical trials.

\end{abstract}

%%%%%%%%%%%%%%%%%%%%%%%%%%%%%%%%%%%%%%%%%%%%%%%%%%%%%%%%%%%%%%%%%%%%%%%%%%%%%%%%
\section{INTRODUCTION}

Respiratory diseases encompass a range of conditions that affect the airways and other structures of the lung. Asthma is one of the most prevalent chronic respiratory conditions, characterized by inflammation and narrowing of the airways that lead to difficulty breathing. According to the World Health Organization (WHO), this condition affected around 262 million people and caused around half a million mortalities in 2019~\cite{Asthma}. Asthma can affect children and adults and is the most common chronic disease among children~\cite{Asthma}. While asthma cannot be cured, its symptoms can be controlled with proper medical care~\cite{asthma_2024}.

\begin{figure}[]
	\centering
		\includegraphics[width=10cm, height=6.5cm, keepaspectratio=true]{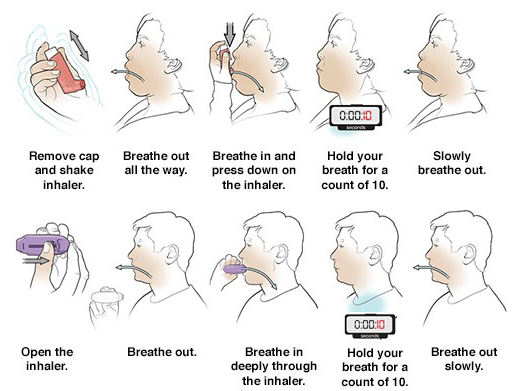}
	  \caption{Correct medication delivery technique by two types of respiratory inhalers. Top: pMDI, Bottom: DPI. Adapted from~\cite{pmdi_inhaler} and ~\cite{dpi_inhaler}, respectively.}\label{fig:inhalers}
\end{figure}

Asthma controller medications play a crucial role in managing this condition by reducing airway inflammation, preventing the symptoms, and minimizing the risk of asthma attacks~\cite{Bateman2008Global}. Such controller medications are typically administered using handheld inhalers. Pressurized metered-dose inhalers (pMDI) and dry powder inhalers (DPI) are the most commonly used devices for managing a wide range of respiratory conditions including but not limited to asthma. Fig. \ref{fig:inhalers} illustrates the correct techniques for administering medication using pMDI and DPI inhalers. To ensure optimal asthma management and treatment outcomes, the proper usage of inhaler devices is of great importance~\cite{Engelkes2015Medication}. It has been shown that several user technique errors can contribute to low adherence to inhaler medication. Fast inhalation, insufficient breath hold, and insufficient inhalation flow rate are only a few examples of user errors~\cite{Sriram2016Suboptimal, Gregoriano2018Use, Bosnic-Anticevich2018Identifying}. Such errors can significantly reduce the effectiveness of medication which could in turn lead to poor symptom control and increased risks of complications and mortalities~\cite{Usmani2018Critical}.

Different actions or events associated with inhaler usage, including actuation, inhalation, and exhalation generate noises called \textit{inhaler sounds}. The automated classification of inhaler sounds has recently been explored as a method for assessing adherence to inhaler technique ~\cite{Ntalianis2020Deep, McNulty2019Automatic,Nousias2016Monitoring,NikosFakotakis2023AI,Pettas2019Recognition}. Inhaler sound analysis can provide valuable insights into medication adherence and usage patterns, helping healthcare professionals monitor patient compliance and adjust treatment plans accordingly. However, a limitation of the existing inhaler sound classification models is that they can only be applied to audio recordings captured from either pMDIs or DPIs. As depicted in Fig. \ref{fig:inhalers}, unlike pMDIs, which rely on patient actuation to deliver the drug, DPIs are activated by the patient's deep and forceful inhalation. As a result, similar events produce different acoustic characteristics in these two classes of inhalers. Consequently, classification models trained on pMDI sounds may not generalize well to DPI sounds, and vice versa. This limitation could reduce the practical applicability of such models in real-world settings, where various inhalers with different acoustic properties are used. Furthermore, existing studies have primarily used sounds captured by microphones embedded in inhalers to develop data-driven models. However, they have not explored setups where consumer devices such as smartwatches are used as audio capture devices. Differences in audio capture hardware introduce another source of variability that can potentially influence the generalizability of inhaler sound classification models. Despite advancements in the development of inhaler sound classification models, the challenges of generalizing across different inhalers and audio capture devices have remained unaddressed.

As mentioned earlier, the previous work used inhalers with embedded microphone sensors to capture and subsequently analyze the inhaler sounds. However, with the advent of consumer devices such as smartwatches and smartphones, there is potential to leverage these technologies for more pervasive monitoring and support. Nowadays smartwatches are generally equipped with high-quality microphone sensors that could be employed for capturing inhaler sounds. Also, smartphones generally benefit from powerful processors that enable complex on-device computations such as training machine learning models. Given the high penetration rate of smart devices, they hold significant potential for monitoring adherence to correct inhaler techniques and supporting patients through coaching to improve their compliance. In this context, this study aims to investigate the possibility of adapting a generic machine learning model to a specific inhaler and smartwatch device using minimal annotated data, which could be implemented on smart devices.

Self-supervised learning (SSL) is a machine learning paradigm where a model learns to generate supervisory signals from the input data itself, without requiring explicit data annotations. SSL techniques utilize unlabeled data to produce meaningful representations which can then be fine-tuned on smaller labeled datasets for specific downstream tasks. wav2vec~2.0 is an audio SSL model developed for speech recognition~\cite{Baevski2020wav2vec}. Since its introduction, this model has successfully been applied to various speech-related tasks~\cite{Fan2021Exploring, Pepino2021Emotion, Xu2021Explore}. It has been shown that the content and quantity of fine-tuned data used to train wav2vec~2.0 influence model performance~\cite{Becerra2022}. There are also examples of applying this model to biomedical data such as brain~\cite{Kostas2021BENDR} and heart sound signals~\cite{Panah2023Exploring}. In this paper, we propose to use the wav2vec~2.0 model for the inhaler sound classification task. The choice of this model is motivated by its state-of-the-art performance on different types of audio signals and its ability to achieve high classification accuracy even when fine-tuned using very small amounts of annotated data~\cite{Baevski2020wav2vec,Panah2023Exploring}. 

By applying wav2vec~2.0 to inhaler sounds, we aim to answer the following research questions:

\textbf{(1) Can wav2vec~2.0 achieve acceptable classification accuracy when trained and tested on DPI inhaler sounds?} To answer this research question, we pre-train and fine-tune the wav2vec~2.0 model on DPI sounds and evaluate its performance on the sounds from the same type of inhaler.

\textbf{(2) Do classification models trained on pMDI sounds generalize to sounds produced by DPI inhalers?} To explore this research question, we evaluate a model pre-trained and fine-tuned on pMDI sounds using DPI sounds.

\textbf{(3) Can we adapt a model trained on pMDI sounds to DPI sounds by re-finetuning it with minimal annotated DPI sounds?} To address this question, we further fine-tune a model already pre-trained and fine-tuned on pMDI sounds using varying amounts of DPI sounds and evaluate its performance on DPI sounds.

The findings indicate that wav2vec~2.0 achieves high classification performance when pre-trained and fine-tuned on DPI inhaler sounds. The results also demonstrate that the proposed model can be adapted to a different inhaler type and audio capture hardware by fine-tuning it on minimal annotated data. To the best of our knowledge, this is the first study in the field demonstrating the feasibility of using smartwatches to capture inhaler sounds and effectively classify them using self-supervised learning techniques to provide a personalized solution for inhaler medication adherence assessment.

\section{Methods}
\begin{figure}[]
	\centering
		\includegraphics[width=8.6cm, height=5.5cm, keepaspectratio=true]{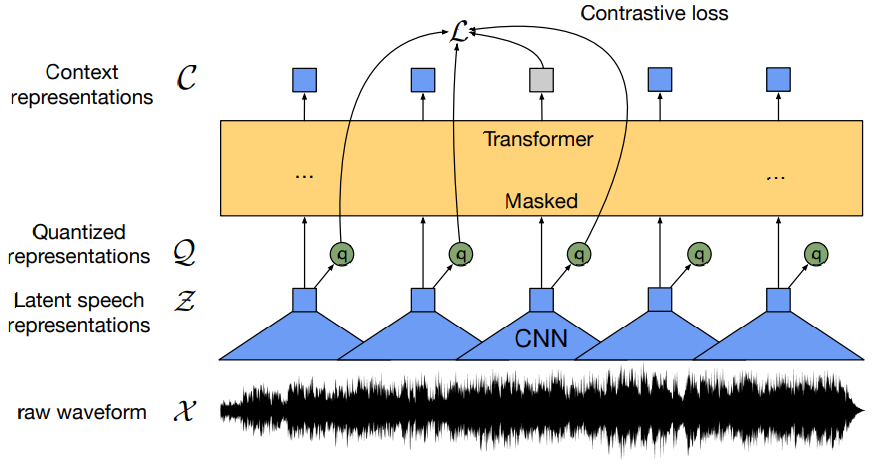}
	  \caption{wav2vec~2.0 model architecture (pre-training phase). Adapted from  \cite{Baevski2020wav2vec}.}\label{fig:w2v}
\end{figure}

\subsection{wav2vec~2.0}
As shown in Fig.  \ref{fig:w2v}, the wav2vec~2.0 model includes three main modules: feature encoder, context network, and quantizer. A summary of each module follows.

\textbf{Feature encoder}: This module comprises a 7-layer convolutional neural network (CNN) that processes the raw audio waveform and extracts high-level features from it.

\textbf{Context network}: This module consists of 12 transformer blocks that process the high-level features from the feature encoder to capture contextual information.

\textbf{Quantizer}: This module discretizes the continuous-valued features obtained from the feature encoder into a sequence of discrete tokens. This discretization step enables more efficient training and inference by reducing the dimensionality of the feature space.

wav2vec~2.0 uses a contrastive loss function for pre-training, encouraging the model to learn discriminative representations of audio signals. The contrastive loss compares the representations of positive and negative samples in a latent space and maximizes the agreement between representations of positive pairs while minimizing the agreement between representations of negative pairs. Readers can refer to~\cite{Baevski2020wav2vec} for more information on wav2vec~2.0 architecture and its training process.

\subsection{Data Pre-processing}\label{sec:data}
We use two inhaler sound datasets for our experiments. The first dataset, DPI-Watch, was collected locally using a Google Pixel Watch 2 as the audio capture device and placebo DPI inhalers (Symbicort Turbuhaler). Ethical approval was obtained from the University College Dublin (UCD) Human Research Ethics Committee. Seven adult participants (six males, and one female), aged 27 to 47, participated in the data collection. The participants were all researchers based at the Quality of Experience Laboratory (QxLab) at UCD. Each participant wore the smartwatch on their wrist throughout the data collection process. The participants were instructed to use the inhaler correctly based on a protocol, as typically performed in a clinical procedure. They then used the placebo inhaler multiple times based on the given instructions and at the same time inhaler sounds were recorded using the smartwatch. There were approximately 10-second pauses between each inhaler use. The data collection was conducted in a quiet room.

The dataset consists of 71 recordings, each lasting between 9 and 28 seconds. Each recording captures the administration of a single dose, which was manually segmented into three classes: actuation (device activation), exhalation, and inhalation, based on auditory inspection. This segmentation resulted in 68 actuation, 123 exhalation, and 71 inhalation segments. Some recordings include patient errors, such as missed activation or exhalation as well as weak or short inhalation. All recordings have a sampling frequency of 44,100 Hz. We use this dataset for both model training and evaluation. The DPI-Watch dataset has been made publicly available \footnote{\url{https://doi.org/10.5281/zenodo.15221786}}

% The second dataset, \textit{DPI-AWatch}, was collected locally using an Apple Watch as the audio capture device and a DPI inhaler (Symbicort Turbuhaler). One adult male subject familiar with the inhaler technique participated in data collection over 2 months. The participant wore the watch on their wrist throughout the data collection process. The dataset contains 20 recordings with durations ranging from 40 to 48 seconds. We segmented the recordings into actuation, exhalation, and inhalation classes through listening. This segmentation resulted in 20 actuation, 67 exhalation, and 36 inhalation segments. All recordings have a sampling frequency of 48,000 Hz. We use this dataset only for model evaluation purposes. We will make this dataset publicly available upon paper acceptance.

The second dataset is the publicly available Respiratory and Drug Actuation (RDA) dataset ~\cite{Nousias2022Respiratory}. It consists of inhaler sound recordings captured in an acoustically controlled environment, free from ambient noise, using a microphone attached to a pMDI inhaler. Three participants (two males, and one female) familiar with inhaler techniques contributed to the data collection.

This dataset contains 360 audio recordings, each 12 seconds long. A human specialist manually annotated and segmented each recording into four classes: drug actuation, exhalation, inhalation, and noise. The dataset includes 193 actuation, 620 exhalation, 319 inhalation, and 505 noise segments with a sampling frequency of 8,000 Hz. Due to the limited number of subjects in this dataset, We use this dataset only for model training. To maintain class consistency with our other dataset, we exclude the segments of \textit{noise} class from the dataset. The remaining segments are then stratified by class and split randomly into 80\% training and 20\% validation sets.

\begin{figure*}[]
	\centering
		\includegraphics[width=16cm, height=16cm, keepaspectratio=true]{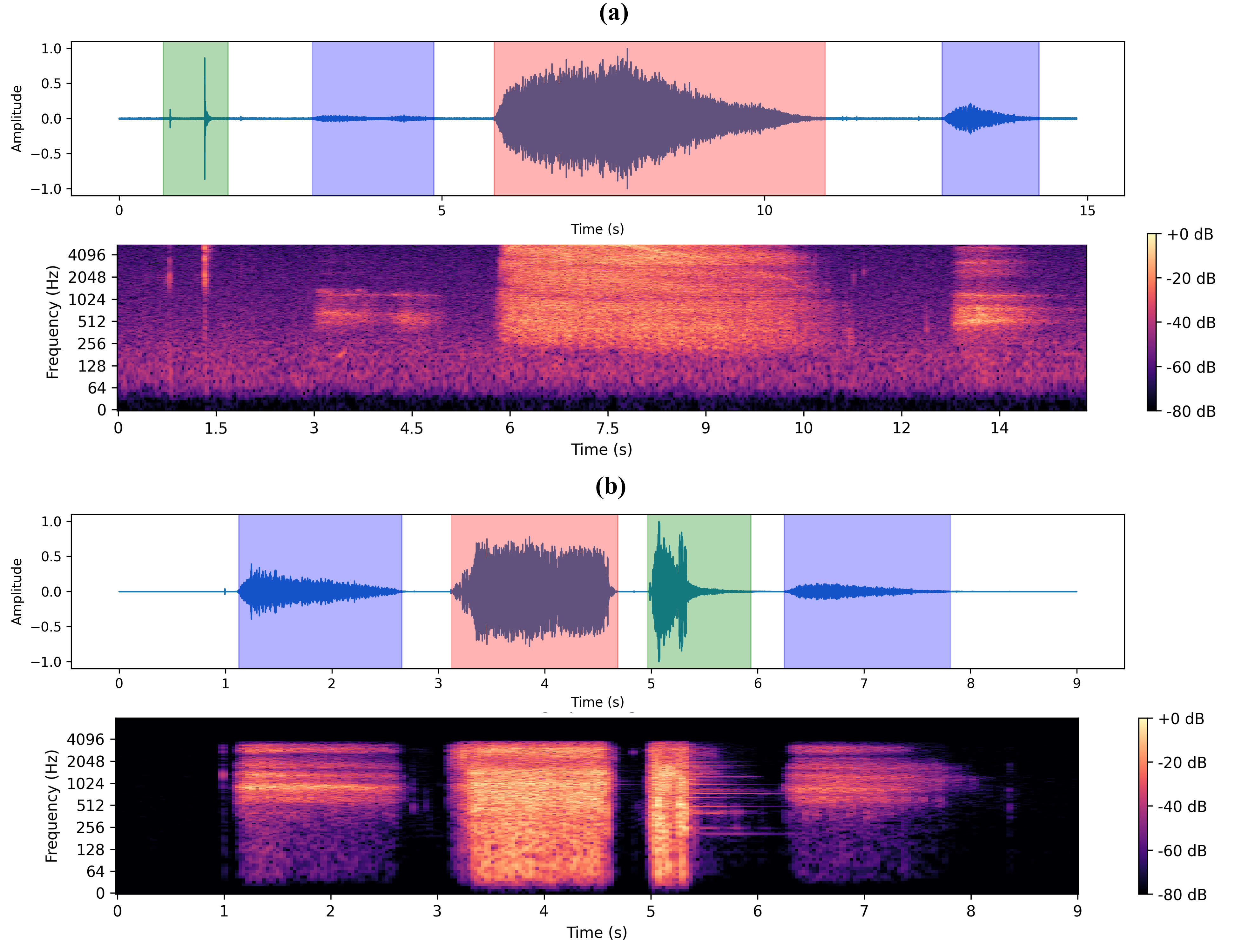}
	  \caption{Examples of sound waveforms colored with ground truth labels, and their corresponding spectrograms captured using (a) DPI and (b) pMDI inhalers. Sound events include actuation (green), exhalation (blue), and inhalation (red). }\label{fig:inhalers_wav_spec}
\end{figure*}

All segments from both datasets are resampled to 16,000 Hz, per the training requirements of the wav2vec~2.0 model~\cite{Baevski2020wav2vec}. Also, amplitude normalization is applied to all segments. Fig. \ref{fig:inhalers_wav_spec} illustrates examples of sound waveforms and their corresponding spectrograms of two recordings from pMDI and DPI inhalers.

\begin{table}[h]
\centering
\renewcommand{\arraystretch}{1.3}
\setlength{\tabcolsep}{8pt}
\caption{Model training configurations}
\label{tb:model-train-configs}
\begin{tabular}{lccc}
\hline
\textbf{Model} & \textbf{\begin{tabular}[c]{@{}c@{}}Pre-training\\Data\end{tabular}} & \textbf{\begin{tabular}[c]{@{}c@{}}Fine-tuning\\Data\end{tabular}} & \textbf{\begin{tabular}[c]{@{}c@{}}Re-finetuning \\Data\end{tabular}} \\ \hline
LS\_DPI & LibriSpeech & DPI-Watch & - \\
DPI\_DPI & DPI-Watch & DPI-Watch & -\\
MDI\_MDI & RDA & RDA & -\\
MDI\_DPI & RDA & RDA & DPI-Watch\\
\hline
\end{tabular}
\end{table}

\subsection{Model Training}\label{sec:model_train}

As summarized in Table~\ref{tb:model-train-configs}, we train multiple classification models with different configurations:

\begin{itemize}
    \item \textbf{LS\_DPI:} We fine-tune the wav2vec 2.0 base model, which was already pre-trained on the LibriSpeech dataset~\cite{panayotov2015librispeech}, on labeled segments from the DPI-Watch dataset.
    
    \item \textbf{DPI\_DPI:} We first pre-train the wav2vec 2.0 model on unlabeled, unsegmented recordings from the DPI-Watch dataset and then fine-tune it on labeled segments from the same dataset.
    
    \item \textbf{MDI\_MDI:} We first pre-train the wav2vec 2.0 model on unlabeled, unsegmented recordings from the RDA dataset and then fine-tune it on labeled segments from the same dataset.
    
    \item \textbf{MDI\_DPI:} We further fine-tune the MDI\_MDI model using labeled segments from the DPI-Watch dataset. Fine-tuning is performed with different amounts of annotated data by gradually reducing the number of segments, from 4.5 minutes of inhaler sounds to less than half a minute. This configuration allows us to investigate the feasibility of adapting a model trained on one type of inhaler to another using minimal amounts of data from the target inhaler.
\end{itemize}

We use the Fairseq framework~\cite{Ott2019fairseq} for model pre-training. To fine-tune the pre-trained models, an average pooling layer and a fully connected layer are added on top of the wav2vec 2.0 model. Fine-tuning of the models is performed for 40 epochs. To avoid overfitting, the training is stopped if validation loss does not decrease for 5 consecutive epochs. The optimization is performed using the AdamW optimizer~\cite{Loshchilov2017Decoupled} and the OneCycle learning rate scheduler~\cite{Smith2019Super-convergence} with a maximum learning rate 3e-5.

\subsection{Model Evaluation}
After the training phase, the models are evaluated using segments from the \textit{DPI-Watch} dataset. Two evaluation methods are employed depending on the model training configuration.

\begin{itemize}
    \item \textbf{{Hold-Out:}} The dataset is split into separate training, validation, and test sets. Data from the first three subjects is used for training, the next three for testing, and the last subject for model evaluation.
    \item \textbf{{One-Subject-Out Cross-Validation (LOSO-CV):}} The model is trained and validated on data from six subjects while being tested on the remaining subject. This process is repeated for each subject, and the results are averaged.
\end{itemize}

Three metrics are used to report the model performance: the first metric is \textit{Recall}, which quantifies the model's performance for each class, including \textit{actuation}, \textit{exhalation} and \textit{inhalation}. The second metric is \textit{unweighted average recall (UAR)}, which measures the model's overall performance. UAR is preferred in our problem as it prevents majority-class dominance and ensures that performance is equally evaluated for each class, regardless of size. This metric is calculated using the following equation:  
\begin{equation}
UAR=\frac{\sum_{i=1}^{N_c} \text {Recall}_i}{N_c}
\end{equation}

\noindent where Recall$_i$ is the recall of the $i$th class, and N$_c$ is the number of classes (three).  

Additionally, we report the \textit{F1-score}, which balances precision and recall for each class and provides an overall measure of model performance. The F1-score is defined as follows:  
\begin{equation}
F1_i = 2 \times \frac{\text{Precision}_i \times \text{Recall}_i}{\text{Precision}_i + \text{Recall}_i}
\end{equation}

\noindent where Precision$_i$ and Recall$_i$ are the precision and recall of the $i$th class, respectively. The overall \textit{macro-averaged F1-score} is then computed as:  
\begin{equation}
F1_{\text{macro}} = \frac{\sum_{i=1}^{N_c} F1_i}{N_c}
\end{equation}

\noindent This ensures that the performance of all classes is weighted equally, making it particularly useful in the presence of class imbalance.

\section {Results and Discussion}
\subsection{DPI Models Performance}\label{sec:scenario1}
This section presents the evaluation results of the LS\_DPI and DPI\_DPI models on the DPI-Watch dataset. Given that these models were trained on DPI-Watch sounds, testing them on the same dataset allows us to assess their generalization to sounds from the same type of inhaler.

\begin{table*}[h]
\centering
\renewcommand{\arraystretch}{1.3}
\setlength{\tabcolsep}{8pt}
\caption{Performance of the LS\_DPI and DPI\_DPI models evaluated on the DPI-Watch dataset. LS\_DPI model is evaluated using both the Hold-Out and LOSO-CV evaluation methods, while the DPI\_DPI model is evaluated using the Hold-Out set only.}
\label{tb:dpi-models-results}
\begin{tabular}{llllllllll}
\hline
\textbf{Model} & \textbf{\begin{tabular}[c]{@{}c@{}}Pre-training\\Data\end{tabular}} & \textbf{\begin{tabular}[c]{@{}c@{}}Fine-tuning\\Data\end{tabular}} & \textbf{\begin{tabular}[c]{@{}c@{}}Evaluation\\Method\end{tabular}} & \textbf{\begin{tabular}[c]{@{}c@{}}Recall\\(Actuation)\end{tabular}} & \textbf{\begin{tabular}[c]{@{}c@{}}Recall\\(Exhalation)\end{tabular}} & \textbf{\begin{tabular}[c]{@{}c@{}}Recall\\(Inhalation)\end{tabular}} & \textbf{UAR} & \textbf{F1-score} \\ \hline
LS\_DPI\_1 & LibriSpeech & DPI-Watch &  Hold-Out  & 1.00 & 0.96 & 0.97 & 0.98 & 0.98 \\
LS\_DPI\_2 & LibriSpeech & DPI-Watch &  LOSO-CV  & $1.00 \pm 0$ & $0.98 \pm 0.03$ & $0.96 \pm 0.07$ & $0.98 \pm 0.03$ & $0.98 \pm 0.03$ \\
DPI\_DPI & DPI-Watch  & DPI-Watch & Hold-Out  & 1.00 & 0.98 & 0.97 & 0.98 & 0.98 \\\hline
\end{tabular}
\end{table*}

% \begin{table*}[h]
% \centering
% \renewcommand{\arraystretch}{1.3}
% \setlength{\tabcolsep}{8pt}
% \caption{Performance of LS\_DPI and DPI\_DPI models evaluated on the DPI-Watch dataset. LS\_DPI model is evaluated using both Hold-Out and LOSO-CV evaluation methods, while DPI\_DPI model is evaluated using the Hold-Out set only.}
% \label{tb:dpi-models-results}
% \begin{tabular}{lccccccccc}
% \hline
% \multicolumn{9}{c}{\textbf{Test Data: DPI-Watch}} \\ \hline
% \textbf{Model} & \textbf{\begin{tabular}[c]{@{}c@{}}Pre-training\\Data\end{tabular}} & \textbf{\begin{tabular}[c]{@{}c@{}}Fine-tuning\\Data\end{tabular}} & \textbf{\begin{tabular}[c]{@{}c@{}}Evaluation\\Method\end{tabular}} & \textbf{\begin{tabular}[c]{@{}c@{}}Recall\\(Actuation)\end{tabular}} & \textbf{\begin{tabular}[c]{@{}c@{}}Recall\\(Exhalation)\end{tabular}} & \textbf{\begin{tabular}[c]{@{}c@{}}Recall\\(Inhalation)\end{tabular}} & \textbf{UAR} & \textbf{F1-score} \\ \hline
% LS\_DPI\_1 & LibriSpeech & DPI-Watch &  Hold-Out  & 1.00 & 0.96 & 0.97 & 0.98 & 0.98 \\
% LS\_DPI\_2 & LibriSpeech & DPI-Watch &  LOSO-CV  & $1.00 \pm 0$ & $0.98 \pm 0.03$ & $0.96 \pm 0.07$ & $0.98 \pm 0.03$ & $0.98 \pm 0.03$ \\
% DPI\_DPI & DPI-Watch  & DPI-Watch & Hold-Out  & 1.00 & 0.98 & 0.97 & 0.98 & 0.98 \\
% \hline
% \multicolumn{9}{c}{\textbf{Test Data: DPI-AWatch}} \\ \hline
% LS\_DPI\_1 & LibriSpeech & DPI-Watch &  Hold-Out  & 1.00 & 0.96 & 0.97 & 0.98 & 0.98 \\
% DPI\_DPI & LibriSpeech & DPI-Watch &  Hold-Out  & 1.00 & 0.96 & 0.97 & 0.98 & 0.98 \\
% \hline
% \end{tabular}
% \end{table*}

As shown in Table \ref{tb:dpi-models-results}, both LS\_DPI and DPI\_DPI models achieve a UAR and F1-score of 0.98. However, at the class level, the DPI\_DPI model performs slightly better on the \textit{exhalation} class than the LS\_DPI model on the hold-out test set. This suggests that pre-training on inhaler sounds can slightly enhance classification performance for the \textit{exhalation} class.

Additionally, the results of the DPI\_DPI model with the LOSO-CV evaluation method are very close to those of the LS\_DPI model with the Hold-Out method. Given the high computational cost of multiple rounds of pre-training on the DPI-Watch dataset, we did not report LOSO-CV results for the DPI\_DPI model. The only study that utilized consumer devices (smartphones) to capture inhaler sounds is that of Eleftheriadou et al. \cite{eleftheriadou2020audio}, in which they achieved an F1-score of 94.85\% for the inhaler sound classification task. However, a direct comparison with our results is not possible because their dataset is not publicly available.

Overall, these results demonstrate that high inhaler sound event classification accuracy can be achieved by pre-training and fine-tuning the wav2vec 2.0 model on the DPI-Watch dataset. In other words, the results indicate that the model generalizes well to inhaler sounds from other patients, provided the test data comes from the same type of inhaler and audio capture hardware.

\subsection{pMDI to DPI Generalization}\label{sec:scenario2}
This section presents the evaluation results of the MDI\_MDI model on the DPI-Watch dataset. As discussed in Section~\ref{sec:model_train}, this model was trained on the RDA dataset (pMDI sounds). As a result, testing the model on the DPI-Watch dataset (DPI sounds) allows us to assess the generalizability of the inhaler sound classification model across different inhaler types and recording hardware. 

\begin{table*}[h]
\centering
\renewcommand{\arraystretch}{1.3}
\setlength{\tabcolsep}{8pt}
\caption{Performance of MDI\_MDI model evaluated on the DPI-Watch dataset (Hold-Out set).}
\label{tb:model-c-results}
\begin{tabular}{cccccccccc}
\hline
\textbf{Model} & \textbf{\begin{tabular}[c]{@{}c@{}}Pre-training\\Data\end{tabular}} & \textbf{\begin{tabular}[c]{@{}c@{}}Fine-tuning\\Data\end{tabular}} & \textbf{\begin{tabular}[c]{@{}c@{}}Recall\\(Actuation)\end{tabular}} & \textbf{\begin{tabular}[c]{@{}c@{}}Recall\\(Exhalation)\end{tabular}} & \textbf{\begin{tabular}[c]{@{}c@{}}Recall\\(Inhalation)\end{tabular}} & \textbf{UAR} & \textbf{F1-score} \\ \hline
MDI\_MDI & RDA & RDA  & 0.52 & 0.68 & 0.42 & 0.54 & 0.55 \\
\hline
\end{tabular}
\end{table*}

As shown in Table \ref{tb:model-c-results}, the overall performance of the MDI\_MDI model is low with a UAR of 0.54. \textit{Exhalation} class shows the highest recall (0.68) across three classes while \textit{inhalation} class achieves the lowest (0.42). These results indicate that the classification model trained on pMDI sounds does not generalize well to sounds produced by a DPI device. Given the significant differences in the acoustic properties of these inhalers and variations in audio capture hardware between train and test datasets, this outcome is expected. Nonetheless, these results serve as a baseline for experiments in the next section.

\subsection{Model Adaptation}\label{sec:scenario3}
This section presents the evaluation results of the MDI\_DPI model on the DPI-Watch dataset. As discussed in Section~\ref{sec:model_train}, the MDI\_DPI model was created by re-finetuning the MDI\_MDI model, which had been pre-trained and fine-tuned on the RDA dataset, using varying amounts of labeled data from the DPI-Watch dataset, ranging from 4.5 minutes (100\% of the available labeled data) to 15 seconds. Each resulting model was then evaluated on the hold-out test set of the DPI-Watch dataset. Table \ref{tb:model-d-results} summarizes the results for this scenario.

\begin{table*}[]
\centering
\renewcommand{\arraystretch}{1.3}
\setlength{\tabcolsep}{8pt}
\caption{Performance of MDI\_DPI model evaluated on the DPI-Watch dataset (Hold-Out set).}
\label{tb:model-d-results}
\begin{tabular}{cccccccc}
\hline
\textbf{Model} & \textbf{\begin{tabular}[c]{@{}c@{}}Fine-tuning\\Data Size\end{tabular}} & \textbf{\begin{tabular}[c]{@{}c@{}}Recall\\(Actuation)\end{tabular}} & \textbf{\begin{tabular}[c]{@{}c@{}}Recall\\(Exhalation)\end{tabular}} & \textbf{\begin{tabular}[c]{@{}c@{}}Recall\\(Inhalation)\end{tabular}} & \textbf{UAR} & \textbf{F1-score} \\ \hline
MDI\_DPI\_1 & 4.5 Min  & 0.97 & 0.96 & 1.0 & 0.98 & 0.98 \\
MDI\_DPI\_2 & 2.0 Min  & 0.97 & 0.93 & 1.0 & 0.97 & 0.96 \\
MDI\_DPI\_3 & 1.0 Min  & 0.87 & 0.95 & 1.0 & 0.94 & 0.94 \\
MDI\_DPI\_4 & 0.5 Min  & 0.68 & 0.95 & 1.0 & 0.87 & 0.88 \\
MDI\_DPI\_5 & 15 Sec  & 0.45 & 0.98 & 0.97 & 0.80 & 0.81 \\
\hline
\end{tabular}
\end{table*}

As shown in Table \ref{tb:model-d-results}, with just 4.5 minutes of fine-tuning data, this model achieves a UAR and F1-score of 0.98. Comparing these results with those in Table \ref{tb:model-c-results} shows that further fine-tuning the MDI\_MDI model on the DPI-Watch dataset significantly improves its performance. Moreover, reducing the fine-tuning data size only slightly impacts the model’s performance for the \textit{exhalation} and \textit{inhalation} classes. Even with just around 15 seconds of inhaler sounds, equivalent to three inhaler uses, the model achieves recall values of 0.98 and 0.97 for the \textit{exhalation} and \textit{inhalation} classes, respectively.

As for the \textit{actuation} class, the model achieves a recall of 0.97 with only 2 minutes of fine-tuning data. However, unlike the exhalation and inhalation classes, reducing the size of the fine-tuning data to less than 2 minutes reduces the model's accuracy for the actuation class. The activation mechanisms differ significantly between pMDI and DPI inhalers, which explains the low recall for the actuation class when very small amounts of fine-tuning data are used. If the train and test sets included sounds from the same class of inhalers (either two different DPIs or two different pMDIs), we would expect to achieve a high recall for the actuation class in such scenarios as well. Also, given that most inhaler usage errors are related to incorrect inhalation or exhalation~\cite{Sriram2016Suboptimal, Gregoriano2018Use, Bosnic-Anticevich2018Identifying}, misclassification of the actuation class is clinically less critical than that of the other two classes.

These results indicate that re-finetuning the wav2vec~2.0 model initially trained on data from one type of inhaler using small amounts of sounds from a target inhaler is a promising approach to adapting the model to a different inhaler and audio capture hardware.

\subsection{Overall Analysis}
In Section \ref{sec:scenario1}, we showed that high classification accuracy can be achieved by pre-training and fine-tuning the wav2vec 2.0 model on DPI sounds. Our analysis in Section \ref{sec:scenario2} confirmed that, as expected, a model trained on pMDI sounds does not perform well when tested on DPI sounds. Several fundamental factors contribute to this poor performance. The first factor is acoustic variability due to differences in inhaler types. DPI inhalers produce distinct acoustic characteristics compared to pMDI inhalers. Variations in sound frequency and intensity between these two inhaler types can impact the model’s ability to generalize. The second factor is audio capture hardware. pMDI sounds were recorded using a microphone attached to the inhaler, whereas DPI sounds were captured by a smartwatch microphone. These microphones may differ in frequency response range, sensitivity, and directionality, which can, in turn, influence the captured audio signals and reduce the model’s generalization ability. The third factor is device positioning. The placement of the smartwatch on the subject’s wrist may introduce acoustic variations compared to a microphone directly attached to the inhaler. Differences in distance, angle, and orientation relative to the inhaler can affect the recorded sounds and further challenge the model’s ability to generalize.

In Section \ref{sec:scenario3}, we demonstrated that a model trained on pMDI sounds can be adapted to DPI sounds through re-finetuning with minimal data from a target inhaler. Specifically, we showed that re-finetuning the model with just 15 seconds of inhaler sounds can lead to high classification accuracy for the clinically important \textit{exhalation} and \textit{inhalation} classes. These results highlight the potential for developing a generic pre-trained and fine-tuned classification model that can be quickly and easily re-finetuned using only a few audio samples from a specific inhaler.

\section{Conclusion}
This paper investigated the application of self-supervised learning, and in particular, the wav2vec~2.0 model for inhaler sound classification. We demonstrated that high classification performance can be achieved by pre-training and fine-tuning this model on inhaler sound data. We also showed the potential to take a generic inhaler sound classification model and adapt it to a target inhaler and audio capture hardware using a relatively small amount of labeled data (e.g., 5 minutes). Additionally, We showed that re-finetuning the model with minimal amounts of data from a target inhaler (e.g., 30 seconds) is a promising approach for personalized respiratory compliance and coaching applications that could be potentially integrated into a mobile phone application connected to a smartwatch. This is the first study illustrating the potential to use consumer devices, particularly smartwatches, for monitoring inhaler use with personalized machine learning models. Our findings pave the way for future research on developing personalized inhaler coaching applications, patient monitoring, and clinical trial data opportunities.

In the future, we plan to extend our work by validating the proposed models on a larger cohort of users and across different inhaler devices. Additionally, we aim to analyze not only classification performance but also the efficacy of inhaler usage, providing deeper insights into patient medication adherence. This will help facilitate the real-world deployment of inhaler monitoring systems, supporting both personalized respiratory coaching and large-scale clinical trials of inhaler-based treatments.

\addtolength{\textheight}{-6cm}   % This command serves to balance the column lengths

%%%%%%%%%%%%%%%%%%%%%%%%%%%%%%%%%%%%%%%%%%%%%%%%%%%%%%%%%%%%%%%%%%%%%%%%%%%%%%%%

%%%%%%%%%%%%%%%%%%%%%%%%%%%%%%%%%%%%%%%%%%%%%%%%%%%%%%%%%%%%%%%%%%%%%%%%%%%%%%%%

%%%%%%%%%%%%%%%%%%%%%%%%%%%%%%%%%%%%%%%%%%%%%%%%%%%%%%%%%%%%%%%%%%%%%%%%%%%%%%%%

\section*{ACKNOWLEDGMENT}

This work was conducted with research grants from the Science Foundation Ireland (SFI) co-funded under the European Regional Development Fund under Grant Numbers 12/RC/2289\_P2 and 13/RC/2077\_P2. For the purpose of Open Access, the author
has applied a CC BY public copyright license to any Author
Accepted Manuscript version arising from this submission.

%%%%%%%%%%%%%%%%%%%%%%%%%%%%%%%%%%%%%%%%%%%%%%%%%%%%%%%%%%%%%%%%%%%%%%%%%%%%%%%%

\end{document}